%% file: paper.tex
\documentclass[conference,letterpaper,twocolumn,final,twoside]{IEEEtran}

\usepackage{multirow}
\usepackage{ifpdf}
\usepackage{color}
\usepackage[latin1]{inputenc}
\usepackage{cite,url}
\usepackage{amssymb}
\usepackage{amsmath}
\usepackage{mathtools}
\usepackage{subcaption}
\usepackage{graphicx}
\usepackage{amsmath}
\usepackage{amssymb}
\usepackage{tikz}
\usepackage{arrayjob}
\usepackage{algorithm2e}
\usepackage[artemisia]{textgreek}
\usepackage{multirow}
\usepackage{dcolumn}

\usepackage{pgfplots}
\pgfplotsset{compat=1.3}% <-- moves axis labels near ticklabels (respects tick label widths)

\usetikzlibrary{external}
\usepgfplotslibrary{external}
\tikzsetexternalprefix{}

\newcommand{\deri}{\mathrm{d}}

\newcommand{\musec}{\text{\textmu s}}

\newcommand{\ies}{\text{IES}}
\newcommand{\iesstar}{\text{IES}^{*}}

\begin{document}

%page numbers on top with chapter heading
\title{Precision of Pulse-Coupled Oscillator Synchronization on FPGA-Based Radios}

\author{\authorblockN{G\"unther Brandner\authorrefmark{1}, Johannes Klinglmayr\authorrefmark{1}, Udo Schilcher\authorrefmark{1}, Dominik Egarter\authorrefmark{1}, and Christian Bettstetter\authorrefmark{1}\authorrefmark{4}}
\vspace{0.2cm}
\authorblockA{\authorrefmark{1}University of Klagenfurt, Mobile Systems Group, Inst. of Networked and Embedded Systems, Klagenfurt, Austria\\
%www: nes.aau.at
}
\authorblockA{\authorrefmark{4}Lakeside Labs GmbH, Klagenfurt, Austria\\[0.2cm]
E-Mail: guenther.brandner@aau.at}}

\sloppy

\maketitle

\thispagestyle{empty}

\input{abstract.tex}
\vspace{0.3cm}

\input{intro.tex}
\vspace{0.3cm}

\input{main.tex}

\vspace{0.3cm}

\input{conclusion.tex}

\bibliographystyle{ieeetr}    %in order of reference
\bibliography{paper,bettstetter}

\end{document}

%% file: abstract.tex
\begin{abstract}
The precision of synchronization algorithms based on the theory of pulse-coupled oscillators is evaluated on FPGA-based radios for the first time. Measurements show that such algorithms can reach precision in the low microsecond range when being implemented in the physical layer. Furthermore, we propose an algorithm extension accounting for phase rate deviations of the hardware and show that an improved precision below one microsecond is possible with this extension in the given setup. The resulting algorithm can thus be applied in ad hoc wireless systems for fully distributed synchronization of transmission slots or sleep cycles, in particular, if centralized synchronization is impossible.
\end{abstract}

%% file: intro.tex
\section{Introduction and Motivation}\label{sec:intro}

The mathematical modeling of pulse-coupled biological oscillators, as proposed in \cite{mirollo1990synchronization} inspired by \cite{Peskin1975}, offers a fully decentralized and scalable approach for time synchronization. There is a broad spectrum of work on pulse-coupled oscillators in physics, biology, neuroscience, and other disciplines (see, e.g., \cite{Buck81,abbott93,Vreeswijk1994,Ernst1995,Gerstner1996,Ernst1998,Golomb01,Nishimura2011,klinglmayr12:njp} and references therein).
The communications engineering community has been interested to transfer these results to self-organizing synchronization of nodes in wireless systems~\cite{mathar1996pulse} for purposes such as slot and frame synchronization, scheduling of cooperative transmissions and sleep cycles, and distributed sensing. A one-to-one transfer is, however, infeasible due to the differences between biological and radio communications. Several extensions and modifications are required with respect to delays, noise, multihop communications, and synchronization words, to mention a few (see \cite{mathar1996pulse,hong2005scalable,Lucarelli2004,Werner-Allen2005,tyrrell06:bionetics,simeone2008distributed,tyrrell10:tmc,6571262,Wang2012} and references therein).

Despite the conceptional and theoretical advances in the design of pulse-coupled oscillator synchronization algorithms for wireless systems, real-world performance studies and experimental proofs of concepts are largely missing. There only exist a few implementations on low-cost wireless sensor platforms (see \cite{Werner-Allen2005,leidenfrost2009,pagliari2011}), whose results are of interest, but their synchronization precision is limited by hardware capabilities. For example, the ``fifty percentile group spread" in a system with $24$ MicaZ motes reported in \cite{Werner-Allen2005} is in the order of $100\,\musec$, which is insufficient for certain applications, such as slot and frame synchronization.

This paper intends to advance this direction of research. In particular, we evaluate the synchronization precision that algorithms based on pulse-coupled oscillators achieve in practice when being integrated into the physical layer on a programmable radio platform. Furthermore, as a result of our experimental research, we gain further insight into the behavior of pulse-coupling in real-world wireless scenarios, and thus propose an extension to the theory, which intends to correct phase rate deviations.

Our main contributions are as follows:
\begin{itemize}
\item Providing a proof of concept by implementing three pulse-coupled oscillator algorithms on field-programmable gate array (FPGA)-based radios
\item Comparing the synchronization precision of these algorithms by measurements
\item Proposing an extension to the synchronization algorithms accounting for phase rate deviations
\item Showing by experiments that precisions below one $\musec$ can be achieved
\end{itemize}
To the best of our knowledge, this is the first lower-layer implementation and real-world performance study of recent pulse-coupled oscillator algorithms on programmable radios.

%% file: main.tex
\section{Synchronization Algorithms} \label{sec:sync_algorithms}

We evaluate the synchronization precision of the following recently proposed synchronization algorithms:
\begin{itemize}
\item Synchronization by Pagliari \& Scaglione (PS)~\cite{pagliari2011,5727895},
\item Synchronization  with inhibitory coupling and self-adjustment (SISA) \cite{klinglmayr12:taas} and
\item Synchronization  with inhibitory and excitatory coupling with stochastic pulse emission ($\ies$)~\cite{klinglmayr12:njp}.
\end{itemize}
We also propose and evaluate a modified version of the IES algorithm that applies phase rate correction ($\iesstar$). The objective of all four algorithms is to synchronize the phases of oscillators.

The general procedure is as follows: The oscillator's phase $\phi$ is increased from zero to one. When $\phi$ reaches one, $\phi$ is reset and a pulse is emitted, either always or with probability $p<1$ depending on the algorithm. When receiving a pulse from another oscillator, an oscillator adjusts its own phase according to an update function $H(\phi)$.

The absolute time is called $t$. The period $\tau_{i\!j}$ denotes the delay between oscillator $i$ and oscillator $j$, i.e., the time it takes from the start of a pulse at $i$ until it is processed at $j$. Let $\tau_{\min}$, $\tau_{\max}$, and $\overline{\tau}$ denote the minimum, maximum, and mean values of all delays, respectively. Furthermore, $\phi(t)$ is an oscillator's phase at time $t$ and $\phi(t^{+})$ its phase infinitely short after $t$. The term $\nu_i$ is the phase rate deviation of oscillator $i$, i.e., the speed of oscillator $i$ compared to a reference oscillator; $\nu_{\max}$ is the maximum phase rate deviation of all oscillators. Let $h(t)$ denote the function which maps durations $t$ in seconds to the corresponding phase, i.e. $h(t)=\frac{t}{t_c}$, where $t_c$ is the cycle length in seconds.

\begin{figure}[!ht]

\begin{subfigure}{0.95\columnwidth}
\fcolorbox{black}{white}{
\begin{minipage}[b]{1\columnwidth}
\begin{enumerate}
\item Whenever $\phi(t)=1$, the oscillator sends a pulse.
\item The refractory interval is
\begin{equation*}
[0,\phi_{\mathrm{ref}}=2 \, (1+\nu_{\max}) \, h(\tau_{\max})].
\end{equation*}
\item Upon reception of a pulse at time $t_{*}$: %=t+\tau_{ij}$:
\begin{eqnarray*}
\phi(t_{*}^{+})=\begin{cases}
\phi(t_{*})  & \text{if $\phi(t_{*}) \leq \phi_{\mathrm{ref}}$,} \\
H_{\text{PS}}(\phi(t_{*})) & \text{else.}
\end{cases}
\end{eqnarray*}
\end{enumerate}
\end{minipage}
}
\caption{PS}
\end{subfigure}% need this comment symbol to avoid overfull hbox

\vspace{0.5cm}

\begin{subfigure}{0.95\columnwidth}
\fcolorbox{black}{white}{
\begin{minipage}[b]{1\columnwidth}
\begin{enumerate}
\item Whenever $\phi(t)=1$, the oscillator adjusts its phase to $\phi(t^{+})=H_{\text{SISA}}(1)$ and sends a pulse.
\item The refractory interval is
\begin{equation*}
[0,\phi_{\mathrm{ref}}=H_{\text{SISA}}(1)+ 2 \, (1+\nu_{\max}) \, h(\tau_{\max})].
\end{equation*}
\item Upon reception of a pulse at time $t_{*}$: %=t+\tau_{ij}$:
\begin{eqnarray*}
\phi(t_{*}^{+})=\begin{cases}
\phi(t_{*})  & \text{if $\phi(t_{*}) \leq \phi_{\mathrm{ref}}$,} \\
H_{\text{SISA}}(\phi(t_{*})) & \text{else.}
\end{cases}
\end{eqnarray*}
\end{enumerate}
\end{minipage}
}
\caption{SISA}
\end{subfigure}% need this comment symbol to avoid overfull hbox

\vspace{0.5cm}

\begin{subfigure}{0.95\columnwidth}
\fcolorbox{black}{white}{
\begin{minipage}[b]{1\columnwidth}
\begin{enumerate}
\item Whenever $\phi(t)=1$, the oscillator sends a pulse with probability $p<1$.
\item The refractory interval is \[[0,\phi_{\mathrm{ref}}=(1+\nu_{\max}) \,h(\tau_{\max})].\]
\item Upon reception of a pulse at time $t_{*}$: %=t+\tau_{ij}$:
\begin{eqnarray*}
\phi(t_{*}^{+})=\begin{cases}
\phi(t_{*})  & \text{if $\phi(t_{*}) \leq \phi_{\mathrm{ref}}$,} \\
H_{\ies}(\phi(t_{*})) & \text{else.}
\end{cases}
\end{eqnarray*}
\end{enumerate}
\end{minipage}
}
\caption{$\ies$}
\end{subfigure}% need this comment symbol to avoid overfull hbox

\vspace{0.5cm}

\begin{subfigure}{0.95\columnwidth}
\fcolorbox{black}{white}{
\begin{minipage}[b]{1\columnwidth}
\begin{enumerate}
\item Whenever $\phi(t)=1$ and no pulse has been received within the last $\overline{\tau}-\tau_{\min}$ seconds, the oscillator sends a pulse with probability $p<1$.
\item The refractory interval is \[[0,\phi_{\mathrm{ref}}=(1+\nu_{\max}) \, h(\tau_{\max})].\]
\item Upon reception of a pulse at time $t_{*}$: %=t+\tau_{ij}$:
\begin{eqnarray*}
\phi(t_{*}^{+})=\begin{cases}
\phi(t_{*})  & \text{if $\phi(t_{*}) \leq \phi_{\mathrm{ref}}$,} \\
H_{\iesstar}(\phi(t_{*})) & \text{else.}
\end{cases}
\end{eqnarray*}
\end{enumerate}
\end{minipage}
}
\caption{$\iesstar$}
\end{subfigure}% need this comment symbol to avoid overfull hbox
\caption{Synchronization algorithms} \label{fig:sync_algorithms}
\vspace{-0.8cm}
\end{figure}
Figure~\ref{fig:sync_algorithms} specifies the four algorithms.
We use
\begin{equation*}
H_{\text{PS}}(\phi)=\min(1,a_1\phi+a_0)
\end{equation*}
for PS with parameters $a_0=1$ and $a_1=\exp(1)$ (strong coupling). These parameters for PS are chosen as they optimize the convergence speed. Choosing different parameters does not influence the achieved precision. For SISA we apply
\begin{equation*}
H_{\text{SISA}}(\phi)=(1+\alpha)\,\phi~\mathrm{mod}~1,
\end{equation*}
with $\alpha=0.5$ which is also applied in \cite{klinglmayr12:taas}.
IES uses
$H_{\ies}(\phi)=\widetilde{H}_{\ies}(\phi-h(\tau_{\min})~\mathrm{mod}~1)+h(\tau_{\min})~\mathrm{mod}~1$.
$\widetilde{H}_{\ies}$ is, \emph{mutatis mutandis}, a function of the form~\cite{Klinglmayr13}
\begin{eqnarray*}
\widetilde{H}_{\ies}(\phi)=
\begin{cases}
%\phi          & \text{if $\phi \leq \phi_{\mathrm{ref}}$},\\
h_1(\phi)     & \text{if $\phi_{\mathrm{ref}} < \phi \leq \frac{1}{2}$},\\
h_2(\phi)     & \text{if $\frac{1}{2} < \phi \leq 1$,}
\end{cases}
\end{eqnarray*}
where $h_1$ and $h_2$ are continuous functions that satisfy the following requirements:
\begin{itemize}
    \item $\frac{\deri h_1}{\deri \phi}>0$, $\frac{\deri h_2}{\deri \phi}>0$,
    \item $h_1(\tau_{\max})=\tau_{\max}$,
    \item $h_1(0.5) \leq 0.25 -\tau_{\max}-\tau_{\min}$,
    \item $h_2(0.5^{+}) \geq 0.75+(\tau_{\max}-\tau_{\min})$, and
    \item $h_2(1)=1$.
\end{itemize}
In the following we use
\[h_1(\phi)=a \cdot [\phi-h(\tau_{\max})] + h(\tau_{\max}) \]
and
\[h_2(\phi)=b \cdot [\phi-1]+1,\]
with
$a=\frac{\frac{1}{4}-2h(\tau_{\max})-h(\tau_{\min})}{\frac{1}{2}-h(\tau_{\max})}$
and
$b=\frac{1}{2}+2h(\tau_{\min})-2h(\tau_{\max})$. These functions fulfill all
requirements. $\iesstar$ uses $H_{\iesstar}(\phi)=\widetilde{H}_{\ies}(\phi-h(\overline{\tau})~\mathrm{mod}~1)+h(\overline{\tau})~\mathrm{mod}~1$.
For both $\ies$ and $\iesstar$ we apply $p=\frac{1}{2}$ as the sending probability.

\section{Implementation on Programmable Radio}

\begin{figure*}[!ht]
\begin{center}
\begin{subfigure}[a]{\textwidth}
\begin{center}
\includegraphics[scale=.7]{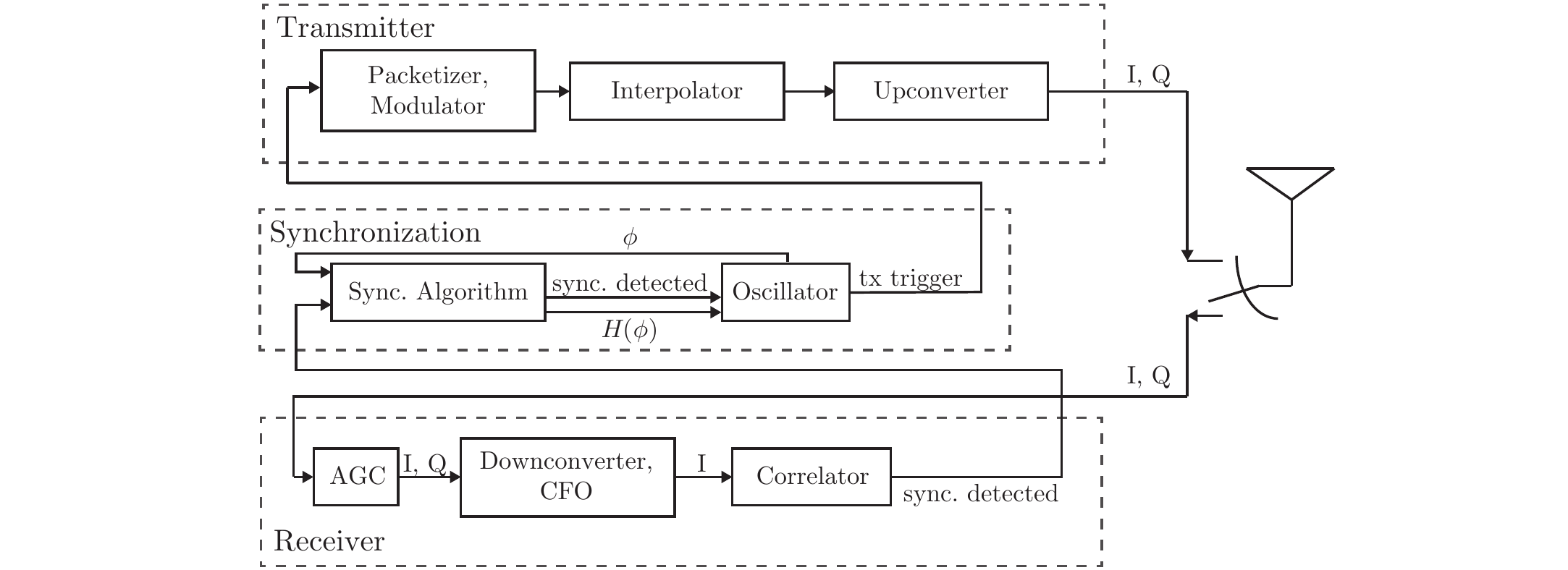}
\caption{Overall structure}
\end{center}
\end{subfigure}

\vspace{0.5cm}
\begin{subfigure}[a]{\textwidth}
\begin{center}
\includegraphics[scale=.7]{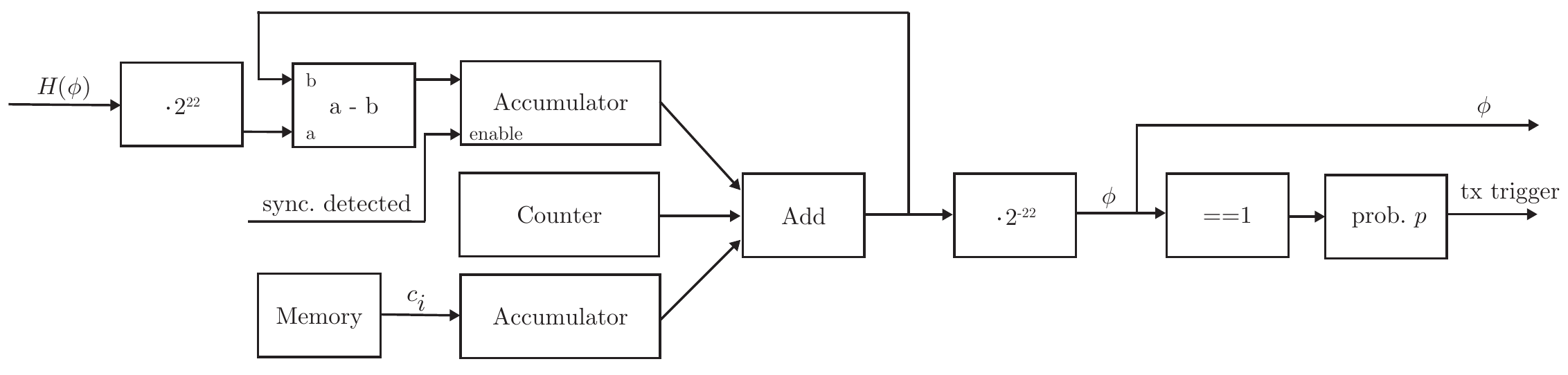}
\caption{Oscillator}
\end{center}
\end{subfigure}
\caption{FPGA design}
\label{fig:fpga_design}
\end{center}
\vspace{-0.8cm}
\end{figure*}

We implement all synchronization algorithms on WARP boards \cite{warpProject}, which are FPGA-based radios. A custom single-carrier physical layer is programmed with $5$~MHz bandwidth and binary phase shift keying (BPSK). Boards operate at $2.4$~GHz and use a peak transmit power of about $20$~dBm. The overall structure of the transceiver is shown in Figure~\ref{fig:fpga_design}(a). All components are implemented directly on the FPGA.
On the transmitter side, the packetizer and modulator build the packet after they receive a trigger signal from the synchronization logic. A modulated packet is fed into an interpolator and upconverter, and finally transmitted over the air. As we cannot send infinitely short pulses, as often assumed in theory, we send short packets instead. These packets have a length of $12$~bytes, where the first $8$~bytes are used for setting the receiving gains of the hardware (agc) and for mitigating carrier frequency offsets (cfo). The remaining $4$~bytes represent a synchronization word consisting of pseudorandom bytes. The transmit duration for a packet is $19.2~\musec$.

On the receiver side, the signals' inphase ($I$) and quadrature ($Q$) components are used to estimate and set the amplifier gains of the boards. The downconverter brings signals to the baseband. We implement a non-data aided algorithm \cite{barry2003} for removing carrier frequency offsets. To detect the synchronization word, a correlator implemented as an FIR filter is applied.

The synchronization logic on the FPGA consists of implementations of the algorithms discussed above. The oscillator component (Figure~\ref{fig:fpga_design}(b)) replicates the oscillator on the board. The main part, generating the phase of the oscillator, is a $22$-bit wrap-around counter running at a clock frequency of $40$~MHz. Thus, the cycle duration $t_c$, i.e., the time it takes for the counter to increment from $0$ to $2^{22}-1$, is about $104.86$~ms. To get a value between zero and one we reinterpret the output as a fractional number by multiplying with $2^{-22}$. After a pulse is detected, the new phase $H(\phi)$, determined by the synchronization algorithm component, is forwarded to the oscillator block and an accumulator is used to adjust to the new phase. Note that the accumulator only processes the value on its input if it is enabled, i.e. if a synchronization word has been detected. The output of the accumulator always reflects the current value, independent on whether or not the accumulator is enabled.

Due to manufacturing tolerances, boards exhibit phase rate deviations. These deviations limit the achievable synchronization precision. As a countermeasure we add correction terms $c_i$ in {$\iesstar$}. These correction terms are determined for each board individually by manually measuring their phase rates with respect to a reference phase. The terms $c_i$ are then stored in the board's memory and applied during synchronization in the following way: at each clock cycle the correction factor $c_i$ is accumulated in a dedicated accumulator and the output of the accumulator is then added to the output of the counter.

Note that phase rates depend on environmental factors, e.g., temperature. The purpose of applying these correction factors is to showcase the influence of phase rate deviations on the achievable synchronization precision. As future work we plan to propose a fully decentralized algorithm that not only synchronizes phases but also phase rates. Note that for all other algorithms, besides  {$\iesstar$}, we set $c_i$ to zero.

\section{Measurement Results}

Six radios are setup to form a fully-connected network with six nodes. The distances between all nodes are a few meters. Thus, all nodes can generally receive packets of all other nodes, but packets might be lost due to, e.g., interference from co-located WLANs.

\subsection{Delays $\tau$ and Phase Rate Deviations $\nu$}

\begin{figure}[!ht]
\begin{center}
\begin{subfigure}[a]{\columnwidth}
\centering
\begin{tikzpicture}[scale=0.8]
\begin{axis}[xlabel={$\tau$ in $\musec$},ylabel={density},legend style={at={(0.99,0.99)},anchor=north east,cells={anchor=west}},grid=major,
xticklabel style={%
/pgf/number format/.cd,
fixed,
fixed zerofill,
precision=1
}]
\addplot [mark repeat=40, mark=none,color=black, line width=1pt]  table [x index=0, y index=1] {./figures/dens_tau.txt};
%\legend{{Warp v1},{Warp v2}}
\end{axis}
\end{tikzpicture}
\caption{Empirical pdf of delay $\tau$}
\end{subfigure}

\vspace{0.5cm}

\begin{subfigure}[a]{\columnwidth}
\centering
\begin{tikzpicture}[scale=0.8]
\begin{axis}[xlabel={time in min},ylabel={$\nu_i$ in ppm},ymin=0,ymax=6.5,,xmax=1000,legend style={at={(0.99,0.99)},anchor=north east,cells={anchor=west}},grid=major,
xticklabel style={%
/pgf/number format/.cd,
fixed,
fixed zerofill,
precision=0
}]
\addplot [mark repeat=40, mark=none,color=black, line width=1pt]  table [x index=0, y index=1] {./figures/phase1.txt};
\addplot [mark repeat=40, mark=none,color=black, line width=1pt]  table [x index=0, y index=1] {./figures/phase2.txt};
\addplot [mark repeat=40, mark=none,color=black, line width=1pt]  table [x index=0, y index=1] {./figures/phase3.txt};
\addplot [mark repeat=40, mark=none,color=black, line width=1pt]  table [x index=0, y index=1] {./figures/phase4.txt};
\addplot [mark repeat=40, mark=none,color=black, line width=1pt]  table [x index=0, y index=1] {./figures/phase5.txt};
\addplot [mark repeat=40, mark=none,color=black, line width=1pt]  table [x index=0, y index=1] {./figures/phase6.txt};
%\legend{{W1},{W2},{W3}, {W4},{W5},{W6}}
\end{axis}
\end{tikzpicture}
\caption{Phase rate deviations $\nu$}
\end{subfigure}
\end{center}
\caption{Empirical pdf of delay $\tau$ and phase rate deviations~$\nu$} \label{fig:tau}
\end{figure}
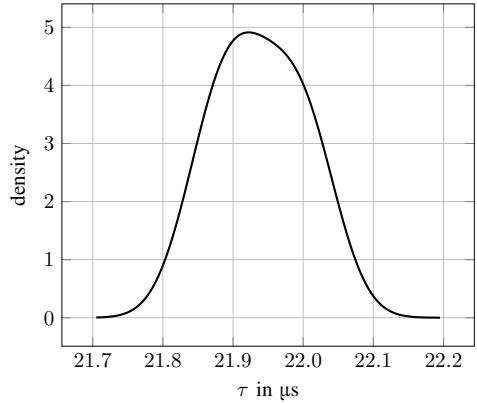
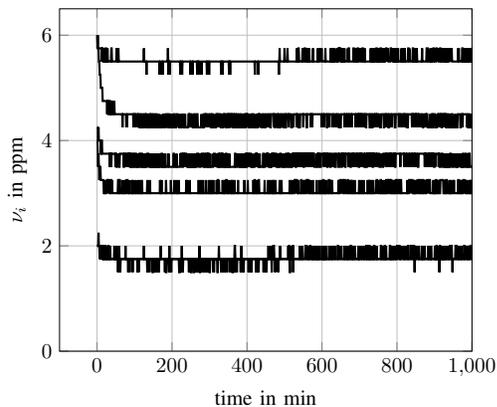

We measure the delays $\tau_{i\!j}$ for various sender-receiver pairs and analyze the overall empirical probability density function (epdf). The epdf is  derived by the method of kernel density estimation~\cite{parzen62}.
Figure~\ref{fig:tau}(a) shows the epdf of $\tau$ based on six sender-receiver pairs and $10\,000$ transmissions each. Experiments show no significant difference between the sender-receiver pairs: The delay is always between $\tau_{\min}=21.7~\musec$ and $\tau_{\max}=22.2~\musec$, and the average delay is $\overline{\tau}=21.92~\musec$; we use these values in the synchronization algorithms. The values presented for $\tau$ are accurate to $\pm 25~$ns.

Figure~\ref{fig:tau}(b) shows the phase rate deviations $\nu$ in parts per million (ppm) from a reference phase for all six nodes. Only five lines are visible as two nodes have the same deviation of about $1.8$ ppm. The phases of all nodes run faster than the reference phase; the deviations range from $1.8$ to $6$ ppm and remain constant over time. The values presented for $\nu$ are accurate to $\pm 0.25~$ppm.

\subsection{Synchronization Precision}

\begin{figure*}[sp]
\begin{center}

\centering
\begin{subfigure}[b]{\textwidth}
\renewcommand\thesubfigure{a}
\begin{subfigure}[b]{0.3\textwidth}
\renewcommand\thesubfigure{i}
\begin{tikzpicture}[scale=0.65]
\begin{axis}[xlabel={cycle},xmin=-5,xmax=180,ymin=0.01,ymax=1000000,ylabel={$\Gamma$ in s},legend style={at={(0.99,0.99)},anchor=north east,cells={anchor=west}},grid=both,
xticklabel style={%
/pgf/number format/.cd,
fixed,
fixed zerofill,
precision=0
},ymode=log,xtick={1,20,40,60,80,100,120,140,160,180,200},
ytick={0.01,0.1,1,10,100,1000,10000,100000,1000000},yticklabels={$10^{-8}$,$10^{-7}$,$10^{-6}$,$10^{-5}$,$10^{-4}$,$10^{-3}$,$10^{-2}$,$10^{-1}$,$1$}]
\addplot [mark repeat=10, mark=o,color=black, line width=1pt]  table [x index=0, y index=1] {./figures/pagliari_prec_n2_qu.txt};
\addplot [mark repeat=10, mark=none,color=black, line width=1pt]  table [x index=0, y index=1] {./figures/pagliari_prec_n2.txt};
\addplot [mark repeat=10, mark=x,color=black, line width=1pt]  table [x index=0, y index=1] {./figures/pagliari_prec_n2_ql.txt};
\legend{$0.95$-quantile, mean, $0.05$-quantile};
\end{axis}
\end{tikzpicture}
\caption{$n=2$}
\end{subfigure}
 \begin{subfigure}[b]{0.3\textwidth}
\renewcommand\thesubfigure{ii}
\begin{tikzpicture}[scale=0.65]
\begin{axis}[xlabel={cycle},xmin=-5,xmax=180,ymin=0.01,ymax=1000000,legend style={at={(0.99,0.99)},anchor=north east,cells={anchor=west}},grid=both,
xticklabel style={%
/pgf/number format/.cd,
fixed,
fixed zerofill,
precision=0
},ymode=log,xtick={1,20,40,60,80,100,120,140,160,180,200},
ytick={0.01,0.1,1,10,100,1000,10000,100000,1000000},yticklabels={$10^{-8}$,$10^{-7}$,$10^{-6}$,$10^{-5}$,$10^{-4}$,$10^{-3}$,$10^{-2}$,$10^{-1}$,$1$}]
\addplot [mark repeat=10, mark=o,color=black, line width=1pt]  table [x index=0, y index=1] {./figures/pagliari_prec_n4_qu.txt};
\addplot [mark repeat=10, mark=none,color=black, line width=1pt]  table [x index=0, y index=1] {./figures/pagliari_prec_n4_mean.txt};
\addplot [mark repeat=10, mark=x,color=black, line width=1pt]  table [x index=0, y index=1] {./figures/pagliari_prec_n4_ql.txt};
\legend{$0.95$-quantile, mean, $0.05$-quantile};
\end{axis}
\end{tikzpicture}
\caption{$n=4$}
\end{subfigure}
 \begin{subfigure}[b]{0.3\textwidth}
\renewcommand\thesubfigure{iii}
\begin{tikzpicture}[scale=0.65]
\begin{axis}[xlabel={cycle},xmin=-5,xmax=180,ymin=0.01,ymax=1000000,legend style={at={(0.99,0.99)},anchor=north east,cells={anchor=west}},grid=both,
xticklabel style={%
/pgf/number format/.cd,
fixed,
fixed zerofill,
precision=0
},ymode=log,xtick={1,20,40,60,80,100,120,140,160,180,200},
ytick={0.01,0.1,1,10,100,1000,10000,100000,1000000},yticklabels={$10^{-8}$,$10^{-7}$,$10^{-6}$,$10^{-5}$,$10^{-4}$,$10^{-3}$,$10^{-2}$,$10^{-1}$,$1$}]
\addplot [mark repeat=10, mark=o,color=black, line width=1pt]  table [x index=0, y index=1] {./figures/pagliari_prec_n6_qu.txt};
\addplot [mark repeat=10, mark=none,color=black, line width=1pt]  table [x index=0, y index=1] {./figures/pagliari_prec_n6_mean.txt};
\addplot [mark repeat=10, mark=x,color=black, line width=1pt]  table [x index=0, y index=1] {./figures/pagliari_prec_n6_ql.txt};
\legend{$0.95$-quantile, mean, $0.05$-quantile};\end{axis}
\end{tikzpicture}
\caption{$n=6$}
\end{subfigure}
\caption{PS}
\end{subfigure}

\vspace{0.5cm}

\begin{subfigure}[b]{\textwidth}
\renewcommand\thesubfigure{b}
\begin{subfigure}[b]{0.3\textwidth}
\renewcommand\thesubfigure{i}
\begin{tikzpicture}[scale=0.65]
\begin{axis}[xlabel={cycle},xmin=-5,xmax=180,ymin=0.01,ymax=1000000,ylabel={$\Gamma$ in s},legend style={at={(0.99,0.99)},anchor=north east,cells={anchor=west}},grid=both,
xticklabel style={%
/pgf/number format/.cd,
fixed,
fixed zerofill,
precision=0
},ymode=log,xtick={1,20,40,60,80,100,120,140,160,180,200},
ytick={0.01,0.1,1,10,100,1000,10000,100000,1000000},yticklabels={$10^{-8}$,$10^{-7}$,$10^{-6}$,$10^{-5}$,$10^{-4}$,$10^{-3}$,$10^{-2}$,$10^{-1}$,$1$}]
\addplot [mark repeat=10, mark=o,color=black, line width=1pt]  table [x index=0, y index=1] {./figures/sisa_prec_n2_qu.txt};
\addplot [mark repeat=10, mark=none,color=black, line width=1pt]  table [x index=0, y index=1] {./figures/sisa_prec_n2_mean.txt};
\addplot [mark repeat=10, mark=x,color=black, line width=1pt]  table [x index=0, y index=1] {./figures/sisa_prec_n2_ql.txt};
\legend{$0.95$-quantile, mean, $0.05$-quantile};
\end{axis}
\end{tikzpicture}
\caption{$n=2$}
\end{subfigure}
\begin{subfigure}[b]{0.3\textwidth}
\renewcommand\thesubfigure{ii}
\begin{tikzpicture}[scale=0.65]
\begin{axis}[xlabel={cycle},xmin=-5,xmax=180,ymin=0.01,ymax=1000000,legend style={at={(0.99,0.99)},anchor=north east,cells={anchor=west}},grid=both,
xticklabel style={%
/pgf/number format/.cd,
fixed,
fixed zerofill,
precision=0
},ymode=log,xtick={1,20,40,60,80,100,120,140,160,180,200},
ytick={0.01,0.1,1,10,100,1000,10000,100000,1000000},yticklabels={$10^{-8}$,$10^{-7}$,$10^{-6}$,$10^{-5}$,$10^{-4}$,$10^{-3}$,$10^{-2}$,$10^{-1}$,$1$}]
\addplot [mark repeat=10, mark=o,color=black, line width=1pt]  table [x index=0, y index=1] {./figures/sisa_prec_n4_qu.txt};
\addplot [mark repeat=10, mark=none,color=black, line width=1pt]  table [x index=0, y index=1] {./figures/sisa_prec_n4_mean.txt};
\addplot [mark repeat=10, mark=x,color=black, line width=1pt]  table [x index=0, y index=1] {./figures/sisa_prec_n4_ql.txt};
\legend{$0.95$-quantile, mean, $0.05$-quantile};
\end{axis}
\end{tikzpicture}
\caption{$n=4$}
\end{subfigure}
\begin{subfigure}[b]{0.3\textwidth}
\renewcommand\thesubfigure{iii}
\begin{tikzpicture}[scale=0.65]
\begin{axis}[xlabel={cycle},xmin=-5,xmax=180,ymin=0.01,ymax=1000000,legend style={at={(0.99,0.99)},anchor=north east,cells={anchor=west}},grid=both,
xticklabel style={%
/pgf/number format/.cd,
fixed,
fixed zerofill,
precision=0
},ymode=log,xtick={1,20,40,60,80,100,120,140,160,180,200},
ytick={0.01,0.1,1,10,100,1000,10000,100000,1000000},yticklabels={$10^{-8}$,$10^{-7}$,$10^{-6}$,$10^{-5}$,$10^{-4}$,$10^{-3}$,$10^{-2}$,$10^{-1}$,$1$}]
\addplot [mark repeat=10, mark=o,color=black, line width=1pt]  table [x index=0, y index=1] {./figures/sisa_prec_n6_qu.txt};
\addplot [mark repeat=10, mark=none,color=black, line width=1pt]  table [x index=0, y index=1] {./figures/sisa_prec_n6_mean.txt};
\addplot [mark repeat=10, mark=x,color=black, line width=1pt]  table [x index=0, y index=1] {./figures/sisa_prec_n6_ql.txt};
\legend{$0.95$-quantile, mean, $0.05$-quantile};
\end{axis}
\end{tikzpicture}
\caption{$n=6$}
\end{subfigure}
\caption{SISA}
\end{subfigure}

\vspace{0.5cm}

\renewcommand\thesubfigure{c}
\begin{subfigure}[a]{\textwidth}
\begin{subfigure}[b]{0.3\textwidth}
\renewcommand\thesubfigure{i}
\begin{tikzpicture}[scale=0.65]
\begin{axis}[xlabel={cycle},xmin=-5,xmax=180,ymin=0.01,ymax=1000000,ylabel={$\Gamma$ in s},legend style={at={(0.99,0.99)},anchor=north east,cells={anchor=west}},grid=both,
xticklabel style={%
/pgf/number format/.cd,
fixed,
fixed zerofill,
precision=0
},ymode=log,xtick={1,20,40,60,80,100,120,140,160,180,200},
ytick={0.01,0.1,1,10,100,1000,10000,100000,1000000},yticklabels={$10^{-8}$,$10^{-7}$,$10^{-6}$,$10^{-5}$,$10^{-4}$,$10^{-3}$,$10^{-2}$,$10^{-1}$,$1$}]
\addplot [mark repeat=10, mark=o,color=black, line width=1pt]  table [x index=0, y index=1] {./figures/ie_prec_n2_qu.txt};
\addplot [mark repeat=10, mark=none,color=black, line width=1pt]  table [x index=0, y index=1] {./figures/ie_prec_n2_mean.txt};
\addplot [mark repeat=10, mark=x,color=black, line width=1pt]  table [x index=0, y index=1] {./figures/ie_prec_n2_ql.txt};
\legend{$0.95$-quantile, mean, $0.05$-quantile};
\end{axis}
\end{tikzpicture}
\caption{$n=2$}
\end{subfigure}
\begin{subfigure}[b]{0.3\textwidth}
\renewcommand\thesubfigure{ii}
\begin{tikzpicture}[scale=0.65]
\begin{axis}[xlabel={cycle},xmin=-5,xmax=180,ymin=0.01,ymax=1000000,legend style={at={(0.99,0.99)},anchor=north east,cells={anchor=west}},grid=both,
xticklabel style={%
/pgf/number format/.cd,
fixed,
fixed zerofill,
precision=0
},ymode=log,xtick={1,20,40,60,80,100,120,140,160,180,200},
ytick={0.01,0.1,1,10,100,1000,10000,100000,1000000},yticklabels={$10^{-8}$,$10^{-7}$,$10^{-6}$,$10^{-5}$,$10^{-4}$,$10^{-3}$,$10^{-2}$,$10^{-1}$,$1$}]
\addplot [mark repeat=10, mark=o,color=black, line width=1pt]  table [x index=0, y index=1] {./figures/ie_prec_n4_qu.txt};
\addplot [mark repeat=10, mark=none,color=black, line width=1pt]  table [x index=0, y index=1] {./figures/ie_prec_n4_mean.txt};
\addplot [mark repeat=10, mark=x,color=black, line width=1pt]  table [x index=0, y index=1] {./figures/ie_prec_n4_ql.txt};
\legend{$0.95$-quantile, mean, $0.05$-quantile};
\end{axis}
\end{tikzpicture}
\caption{$n=4$}
\end{subfigure}
\begin{subfigure}[b]{0.3\textwidth}
\renewcommand\thesubfigure{iii}
\begin{tikzpicture}[scale=0.65]
\begin{axis}[xlabel={cycle},xmin=-5,xmax=180,ymin=0.01,ymax=1000000,legend style={at={(0.99,0.99)},anchor=north east,cells={anchor=west}},grid=both,
xticklabel style={%
/pgf/number format/.cd,
fixed,
fixed zerofill,
precision=0
},ymode=log,xtick={1,20,40,60,80,100,120,140,160,180,200},
ytick={0.01,0.1,1,10,100,1000,10000,100000,1000000},yticklabels={$10^{-8}$,$10^{-7}$,$10^{-6}$,$10^{-5}$,$10^{-4}$,$10^{-3}$,$10^{-2}$,$10^{-1}$,$1$}]
\addplot [mark repeat=10, mark=o,color=black, line width=1pt]  table [x index=0, y index=1] {./figures/ie_prec_n6_qu.txt};
\addplot [mark repeat=10, mark=none,color=black, line width=1pt]  table [x index=0, y index=1] {./figures/ie_prec_n6_mean.txt};
\addplot [mark repeat=10, mark=x,color=black, line width=1pt]  table [x index=0, y index=1] {./figures/ie_prec_n6_ql.txt};
\legend{$0.95$-quantile, mean, $0.05$-quantile};
\end{axis}
\end{tikzpicture}
\caption{$n=6$}
\end{subfigure}
\caption{IES}
\end{subfigure}

\vspace{0.5cm}

\begin{subfigure}[b]{\textwidth}
\renewcommand\thesubfigure{d}
\begin{subfigure}[b]{0.3\textwidth}
\renewcommand\thesubfigure{i}
\begin{tikzpicture}[scale=0.65]
\begin{axis}[xlabel={cycle},xmin=-5,xmax=180,ymin=0.01,ymax=1000000,ylabel={$\Gamma$ in s},legend style={at={(0.99,0.99)},anchor=north east,cells={anchor=west}},grid=both,
xticklabel style={%
/pgf/number format/.cd,
fixed,
fixed zerofill,
precision=0
},ymode=log,xtick={1,20,40,60,80,100,120,140,160,180,200},
ytick={0.01,0.1,1,10,100,1000,10000,100000,1000000},yticklabels={$10^{-8}$,$10^{-7}$,$10^{-6}$,$10^{-5}$,$10^{-4}$,$10^{-3}$,$10^{-2}$,$10^{-1}$,$1$}]
\addplot [mark repeat=10, mark=o,color=black, line width=1pt]  table [x index=0, y index=1] {./figures/ie_prec_n2_adapt_qu.txt};
\addplot [mark repeat=10, mark=none,color=black, line width=1pt]  table [x index=0, y index=1] {./figures/ie_prec_n2_adapt_mean.txt};
\addplot [mark repeat=10, mark=x,color=black, line width=1pt]  table [x index=0, y index=1] {./figures/ie_prec_n2_adapt_ql.txt};
\legend{$0.95$-quantile, mean, $0.05$-quantile};
\end{axis}
\end{tikzpicture}
\caption{$n=2$}
\end{subfigure}
\begin{subfigure}[b]{0.3\textwidth}
\renewcommand\thesubfigure{ii}
\begin{tikzpicture}[scale=0.65]
\begin{axis}[xlabel={cycle},xmin=-5,xmax=180,ymin=0.01,ymax=1000000,legend style={at={(0.99,0.99)},anchor=north east,cells={anchor=west}},grid=both,
xticklabel style={%
/pgf/number format/.cd,
fixed,
fixed zerofill,
precision=0
},ymode=log,xtick={1,20,40,60,80,100,120,140,160,180,200},
ytick={0.01,0.1,1,10,100,1000,10000,100000,1000000},yticklabels={$10^{-8}$,$10^{-7}$,$10^{-6}$,$10^{-5}$,$10^{-4}$,$10^{-3}$,$10^{-2}$,$10^{-1}$,$1$}]
\addplot [mark repeat=10, mark=o,color=black, line width=1pt]  table [x index=0, y index=1] {./figures/ie_prec_n4_adapt_qu.txt};
\addplot [mark repeat=10, mark=none,color=black, line width=1pt]  table [x index=0, y index=1] {./figures/ie_prec_n4_adapt_mean.txt};
\addplot [mark repeat=10, mark=x,color=black, line width=1pt]  table [x index=0, y index=1] {./figures/ie_prec_n4_adapt_ql.txt};
\legend{$0.95$-quantile, mean, $0.05$-quantile};
\end{axis}
\end{tikzpicture}
\caption{$n=4$}
\end{subfigure}
\begin{subfigure}[b]{0.3\textwidth}
\renewcommand\thesubfigure{iii}
\begin{tikzpicture}[scale=0.65]
\begin{axis}[xlabel={cycle},xmin=-5,xmax=180,ymin=0.01,ymax=1000000,legend style={at={(0.99,0.99)},anchor=north east,cells={anchor=west}},grid=both,
xticklabel style={%
/pgf/number format/.cd,
fixed,
fixed zerofill,
precision=0
},ymode=log,xtick={1,20,40,60,80,100,120,140,160,180,200},
ytick={0.01,0.1,1,10,100,1000,10000,100000,1000000},yticklabels={$10^{-8}$,$10^{-7}$,$10^{-6}$,$10^{-5}$,$10^{-4}$,$10^{-3}$,$10^{-2}$,$10^{-1}$,$1$}]
\addplot [mark repeat=10, mark=o,color=black, line width=1pt]  table [x index=0, y index=1] {./figures/ie_prec_n6_adapt_qu.txt};
\addplot [mark repeat=10, mark=none,color=black, line width=1pt]  table [x index=0, y index=1] {./figures/ie_prec_n6_adapt_mean.txt};
\addplot [mark repeat=10, mark=x,color=black, line width=1pt]  table [x index=0, y index=1] {./figures/ie_prec_n6_adapt_ql.txt};
\legend{$0.95$-quantile, mean, $0.05$-quantile};\end{axis}
\end{tikzpicture}
\caption{$n=6$}
\end{subfigure}
\caption{$\iesstar$}
\end{subfigure}
\caption{Synchronization precision} \label{fig:synchronization_precision}
\end{center}
\end{figure*}

The synchronization precision at time $t$ in terms of the maximum phase difference between all nodes is evaluated as~\cite{klinglmayr12:njp}
\begin{eqnarray}
\Gamma(t)= t_c \cdot \max_{i,j}\bigg\{\min[|\phi_i(t)-\phi_j(t)|,1-|\phi_i(t)-\phi_j(t)|]\bigg\}.\nonumber
\end{eqnarray}
Figure~\ref{fig:synchronization_precision} shows the synchronization precision for $n \in \{2,4,6\}$ nodes. Results are based on $100$ synchronization runs, where prior to each run the phases of all nodes are randomly initiated. The $x$-axes show the cycle number. The cycle duration is $52.43$ ms for SISA and $104.86$ ms for all other algorithms. The shown values are accurate to $\pm 25~$ns.

These measurement results can be interpreted as follows: PS converges very quickly to a synchronization precision of about $21$\,$\musec$. For SISA, the speed of convergence decreases with increasing $n$, which is due to the fact that packets lost over the wireless link can cause a deterioration of the synchronization precision: Whenever a node detects a synchronization word from some other node, and is not currently in refractory, it halves its current phase. In case that not all nodes detect the synchronization word, we have the situation that some nodes halve their phases, while others do not halve their phases. The convergence of IES is slower than that of PS with the given parameters, however, it converges to a mean precision of about $1.5$\,$\musec$ ($n=2$), $2$\,$\musec$ ($4$) and $4$\,$\musec$ ($6$). $\iesstar$ achieves a precision of about $200$\,ns ($n=2$), $400$\,ns ($4$) and $600$\,ns ($6$).

The fact that PS and SISA synchronize less precisely than IES in this setup is likely due to two reasons: (i) propagation delays are not considered in those algorithms and (ii) nodes cannot hear other nodes when sending. This result confirms that stochastic communication of synchronization words is an important design feature (see, e.g., \cite{klinglmayr12:njp}). This feature could in principle also be applied to PS.

%% file: conclusion.tex
\section{Conclusions}\label{sec:conclusion}

Measurement results of pulse-coupled oscillator synchronization implemented on FPGA-based radios show that the synchronization precision can reach values below one $\musec$. Key factors for reaching this precision are the explicit consideration of propagation and processing delays, the stochastic nature in communications of synchronization words, and a phase rate correction. The latter mitigates precision limitations caused by phase rate deviations of the hardware.

\section*{Acknowledgements}

This work was supported by Lakeside Labs with funding from ERDF, KWF, BABEG under grant 20214/23794/35530. The work of Johannes Klinglmayr was partially supported by the Linz Center of Mechatronics (LCM) in the framework of the Austrian COMET-K2 program. 

%% file: paper.bbl
\begin{thebibliography}{10}

\bibitem{mirollo1990synchronization}
R.~E. Mirollo and S.~H. Strogatz, ``Synchronization of pulse-coupled biological
  oscillators,'' {\em SIAM J. Appl. Math.}, vol.~50, no.~6, pp.~1645--1662,
  1990.

\bibitem{Peskin1975}
C.~S. Peskin, {\em Mathematical Aspects of Heart Physiology}, pp.~268-- 278.
\newblock Courant Institute of Mathematical Sciences, 1975.

\bibitem{Buck81}
J.~Buck, E.~Buck, J.~Case, and F.~Hanson, ``Control of flashing in fireflies.
  {V. P}acemaker synchronization in {\it pteroptyx cribellata},'' {\em J. Comp.
  Physiology A}, vol.~144, pp.~630--633, Sept. 1981.

\bibitem{abbott93}
L.~F. Abbott and C.~A. van Vreeswijk, ``Asynchronous states in neural networks
  of pulse-coupled oscillators,'' {\em Phys. Rev. E}, vol.~48, pp.~1483--1490,
  1993.

\bibitem{Vreeswijk1994}
C.~A. van Vreeswijk, L.~F. Abbott, and G.~B. Ermentrout, ``When inhibition, not
  excitation, synchronizes neural firing,'' {\em J. Comp. Neurosci.}, vol.~1,
  pp.~313--321, Dec. 1994.

\bibitem{Ernst1995}
U.~Ernst, K.~Pawelzik, and T.~Geisel, ``Synchronization induced by temporal
  delays in pulse-coupled oscillators,'' {\em Phys. Rev. Lett.}, vol.~74,
  pp.~1570--1573, Feb. 1995.

\bibitem{Gerstner1996}
W.~Gerstner, ``Rapid phase locking in systems of pulse-coupled oscillators with
  delays,'' {\em Phys. Rev. Lett.}, vol.~76, pp.~1755--1758, Mar. 1996.

\bibitem{Ernst1998}
U.~Ernst, K.~Pawelzik, and T.~Geisel, ``Delay-induced multistable
  synchronization of biological oscillators,'' {\em Phys. Rev. E}, vol.~57,
  pp.~2150--2162, Feb. 1998.

\bibitem{Golomb01}
D.~Golomb and G.~B. Ermentrout, ``Bistability in pulse propagation in networks
  of excitatory and inhibitory populations,'' {\em Phys. Rev. Lett.}, vol.~86,
  pp.~4179--4182, Apr. 2001.

\bibitem{Nishimura2011}
J.~Nishimura and E.~J. Friedman, ``Robust convergence in pulse-coupled
  oscillators with delays,'' {\em Phys. Rev. Lett.}, vol.~106, May 2011.
\newblock 194101.

\bibitem{klinglmayr12:njp}
J.~Klinglmayr, C.~Kirst, C.~Bettstetter, and M.~Timme, ``Guaranteeing global
  synchronization in networks with stochastic interactions,'' {\em New J.
  Phys.}, vol.~14, July 2012.
\newblock 073031.

\bibitem{mathar1996pulse}
R.~Mathar and J.~Mattfeldt, ``Pulse-coupled decentral synchronization,'' {\em
  SIAM J. Appl. Math.}, vol.~56, no.~4, pp.~1094--1106, 1996.

\bibitem{hong2005scalable}
Y.-W. Hong and A.~Scaglione, ``A scalable synchronization protocol for large
  scale sensor networks and its applications,'' {\em IEEE J. Sel. Areas
  Commun.}, vol.~23, no.~5, pp.~1085--1099, 2005.

\bibitem{Lucarelli2004}
D.~Lucarelli and I.-J. Wang, ``Decentralized synchronization protocols with
  nearest neighbor communication,'' in {\em Proc. ACM Conf. Embedded Networked
  Sensor Systems (SenSys)}, (Baltimore, MD, USA), Nov. 2004.

\bibitem{Werner-Allen2005}
G.~Werner-Allen, G.~Tewari, A.~Patel, M.~Welsh, and R.~Nagpal,
  ``Firefly-inspired sensor network synchronicity with realistic radio
  effects,'' in {\em Proc. ACM Conf. Embedded Networked Sensor Systems
  (SenSys)}, (San Diego, CA, USA), pp.~142--153, Nov. 2005.

\bibitem{tyrrell06:bionetics}
A.~Tyrrell, G.~Auer, and C.~Bettstetter, ``Fireflies as role models for
  synchronization in ad hoc networks,'' in {\em Proc. Intern. Conf. on
  Bio-Inspired Models of Network, Information, and Computing Systems
  (BIONETICS)}, (Cavalese, Italy), Dec. 2006.

\bibitem{simeone2008distributed}
O.~Simeone, U.~Spagnolini, Y.~Bar-Ness, and S.~H. Strogatz, ``Distributed
  synchronization in wireless networks,'' {\em IEEE Signal Process. Mag.},
  vol.~25, no.~5, pp.~81--97, 2008.

\bibitem{tyrrell10:tmc}
A.~Tyrrell, G.~Auer, and C.~Bettstetter, ``Emergent slot synchronization in
  wireless networks,'' {\em {IEEE} Trans.~Mobile Comput.}, vol.~9,
  pp.~719--732, May 2010.

\bibitem{6571262}
Y.~Wang, F.~N\'{u}\~{n}ez, and F.~J. Doyle~III, ``Statistical analysis of the
  pulse-coupled synchronization strategy for wireless sensor networks,'' {\em
  IEEE Trans. Signal Process.}, vol.~61, pp.~5193--5204, Nov. 2013.

\bibitem{Wang2012}
Y.~Wang, F.~N\'{u}\~{n}ez, and F.~J. Doyle~III, ``Increasing sync rate of
  pulse-coupled oscillators via phase response function design: Theory and
  application to wireless networks,'' {\em IEEE Trans. Control Syst. Technol.},
  vol.~21, pp.~1455--1462, July 2013.

\bibitem{leidenfrost2009}
R.~Leidenfrost and W.~Elmenreich, ``Firefly clock synchronization in an
  802.15.4 wireless network,'' {\em EURASIP J. Embed. Syst.}, vol.~2009, 2009.
\newblock 186406.

\bibitem{pagliari2011}
R.~Pagliari and A.~Scaglione, ``Scalable network synchronization with
  pulse-coupled oscillators,'' {\em IEEE Trans. Mobile Comput.}, vol.~10,
  no.~3, pp.~392--405, 2011.

\bibitem{5727895}
R.~Pagliari and A.~Scaglione, ``Correction to 'scalable network synchronization
  with pulse-coupled oscillators','' {\em IEEE Trans. Mobile Comput.}, vol.~10,
  no.~5, 2011.

\bibitem{klinglmayr12:taas}
J.~Klinglmayr and C.~Bettstetter, ``Self-organizing synchronization with
  inhibitory-coupled oscillators: Convergence and robustness,'' {\em ACM
  Trans.~Auton. Adapt. Syst.}, vol.~7, pp.~30:1--30:23, Oct. 2012.

\bibitem{Klinglmayr13}
J.~Klinglmayr, {\em {Self-organizing network synchronization: Convergence and
  robustness in pulse-coupled oscillator systems}}.
\newblock PhD thesis, Alpen-Adria-Universität, Klagenfurt, Austria, 2013.

\bibitem{warpProject}
``{WARP Project: Wireless Open-Access Research Platform, Rice University}.''
\newblock \url{http://warp.rice.edu}; accessed December 14, 2014.

\bibitem{barry2003}
J.~R. Barry, E.~A. Lee, and D.~G. Messerschmitt, {\em Digital Communications}.
\newblock Springer, 2004.

\bibitem{parzen62}
E.~Parzen, ``On estimation of a probability density function and mode,'' {\em
  Ann. Math. Statist.}, vol.~33, no.~3, pp.~1065--1076, 1962.

\end{thebibliography}
